\documentstyle[sprocl,pstricks]{article}

\input{psfig}
\def\Journal#1#2#3#4{{\em #1} {\bf #2}, #3 (#4)}

\begin{document}

\title{DIFFUSIVITY OF Ga AND Al ADATOMS  ON GaAs(001)}
 
\author{A. KLEY AND M. SCHEFFLER}

\address{ Fritz-Haber-Institut, Faradayweg 4-6, D-14195 Berlin-Dahlem}

\maketitle

\abstracts{The diffusivity of  Ga and Al adatoms on the  (2$\times$4) reconstructed
GaAs(001) surface are evaluated using  detailed 
{\it ab initio} total energy calculations of the potential energy surface together
with  transition state theory.
A strong diffusion anisotropy is found, with the direction of fastest diffusion being  parallel to the
surface As-dimer orientation. In contrast to previous calculations we identify 
 a short--bridge position between the two As atoms of a surface dimer as the  adsorption 
site for Al and Ga adatoms.}

\section{Introduction}

Heteroepitaxy of GaAs and AlAs on (001) oriented substrates  
has been successfully applied to create structures of reduced dimensionality.
In spite of the  considerable progress in experimental studies
the knowledge  of the  underlying microscopic  processes  is rather poor.
For example, even for   a key process  like the Ga adatom  diffusion  on  GaAs(001), the 
experimentally derived values for the activation energy~\cite{nishinaga:88,hove:89}
scatter significantly (between  1.1 and 4.0\,eV) and  the  very basic  question ``which is
 the direction
of fastest diffusion on GaAs(001)?''  has not yet  been answered, unambiguously.
A  similar lack of understanding exists with respect to the question of  how the diffusion 
properties of Al and Ga adatoms differ.
This is crucial for controlling
the  growth of high  quality  GaAs/AlAs quantum devices.

In this paper we present calculations of the diffusivities
(activation energies as well as pre--exponential factors)
 of Ga and Al adatoms
on the flat GaAs(001) surface  using  density functional  total energy calculations.

\section{Method}
  Typically the rate constant for 
   adatom hops between neighboring sites is by orders of magnitude smaller
  than the highest phonon frequencies. 
Therefore, an {\it ab initio} molecular dynamics simulation is not a practicable  approach to evaluate
the diffusivity.
It is clearly more appropriate 
to evaluate the Born-Oppenheimer
potential energy surface (PES) and to analyse its properties.
Our study of the diffusivity proceeds in three steps.
At first we determine the PES using density functional theory and the generalized gradient 
approximation~\cite{perdew:92}
for the exchange and correlation potential. This calculation identifies 
the local minima $\{\bf A_{\it I} \}$ of the PES. We then apply transition state theory~\cite{glasstone:41}
(TST) to determine the individual jump rates $\Gamma_{I,J}$ between different sites 
${\bf A_{\it I}}$ and ${\bf A_{\it J}}$. In a third step we use a random walk model 
which assumes uncorrelated 
jumps between the sites $\{\bf A_{\it I} \}$ to obtain the diffusivities in the two orthogonal directions,
[110] and [$\overline{\rm 1}10$]. These  are  the main axes of the diffusion tensor
 according to the symmetry   of the considered  surface structure.

In TST  the jump rate between two  sites ${\bf A_{\it I}}$ and ${\bf A_{\it J}}$ is 
\begin{eqnarray}
\Gamma_{I,J} &  = & \Gamma_{I,J}^{0}\; exp(-\Delta E/k_{\rm B} T)\quad ,
\end{eqnarray}
where $\Delta E$ is the difference between the   energies  of the  adatom at the site  ${\bf A_{\it I}}$
and  at the saddle point  between the sites ${\bf A_{\it I}}$ and ${\bf A_{\it J}}$. 
To find these  sites and the energy  barrier  $\Delta E$,  we calculate  
the  PES for a  diffusing adatom: 
\begin{eqnarray}\label{pes:definition}
E^{\rm PES}(X_{\rm 1},Y_{\rm 1}) = \min_{Z_{\rm 1},\left\{\bf R_{\it i}\right\}_{{\it i}{\rm =2,...N}}} 
E (\{{\bf R}_{\it i}\}) \quad . 
\end{eqnarray}
The set $\{\bf R_{\it i}\}$ describes  the positions of the atoms 
where  $X_{1}$, $Y_{1}$ and $Z_{1}$  are 
the coordinates
of the adatom parallel and perpendicular  the surface, respectively.
$E(\{ {\bf R_{\it i}} \})$ is 
the total energy of the  surface with an adatom. 

The pre--exponential factor $\Gamma_{I,J}^{0}$  is determined within the
harmonic approximation of TST~\cite{vineyard:57}:   
\begin{eqnarray}\label{eq:vorfaktor}
 \Gamma_{I,J}^{0} & = & \frac{\rm 1}{\sqrt{4 \pi^{\rm 2} m_{\rm ad}}} \sqrt{\frac{ det |K_{I}|}{ det | K_{S}' |}}\quad .
\end{eqnarray}
$K_{I}$ and  $K_{S}'$ are the force constant matrices
 for the adatom at the site ${\bf A}_{\it I}$ and at the saddle point 
 between ${\bf A_{\it I}}$ and ${\bf A_{\it J}}$. 
Both  matrices contain  the  force constants  for   ${\bf R_{\rm 1}}$
and for the coordinates ${\bf R_{\it i}}$ of substrate atoms next to the adatom.
According to the TST,  $K_{S}'$ does not contain the mode  perpendicular to the transition plane
between $I$ and $J$.

\section{Results}
We assume in our study that the surface is reconstructed with the 
(2$\times$4) $\beta$2 structure which   consists of  rows of two 
As dimers and two missing dimers in the topmost layer.
This structure  is found to be   present under typical growth conditions~\cite{hashizume:94}
and is also a  structure of lowest surface energy.~\cite{northrup:94}

Our calculations show that  for both adatom species  Al and Ga
there exist two sheets of the   PES  which are  relevant for  diffusion.
The occurrence of several sheets is not too surprising because in the  combined configuration space 
of the adatom and the substrate  
there may be several saddle points between  two  local minima 
${\bf A_{\it I}}$ and ${\bf A_{\it J}}$.
Accordingly the PES as defined in  Eq.~(\ref{pes:definition}) does not
need to be a  unique function.

The two sheets of the PES   differ in the way the adatom interacts with the  surface As-dimers.
Either the surface displays only small relaxations (in particular the dimer bonds remain intact)
when the adatom diffuses across the surface or
the adatom breaks a  surface dimer and causes a strong relaxation. 
For most positions $(X_{\rm 1},Y_{\rm 1})$ of the adatom the two PES are identical. The differences occur
only around the the surface dimers and are caused by an energy barrier to break the
dimers. 

\begin{figure}[htb]
\psfig{figure=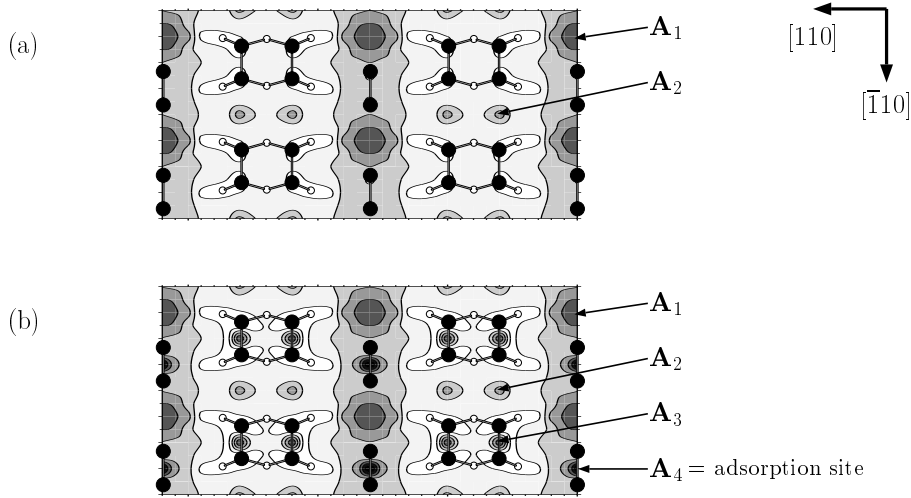,height=6.5cm}
\caption{Potential energy surfaces for a Ga adatom on the GaAs(001)--(2$\times$4) $\beta$2
         surface. (a)~PES sheet obtained when the As surface dimers remain intact. (b)~PES
         sheet obtained when the surface dimers are broken by the adatom. 
         The contour--line spacing is 0.3\,eV.
         The atomic positions 
         of the clean surface are indicated for  atoms  
         of the upper two layers and for the  As-dimers in the third layer (As: filled circles,
         Ga:~empty~circles).
\label{fig:pes}}
\end{figure}

If the adatom does not break the surface dimers (see Fig.\ \ref{fig:pes}(a))  the long--bridge
position  ${\bf A_{\rm 1}}$ between surface dimers 
is  the  favoured  site for adsorption. This  agrees with previous calculations~\cite{ohno:94}
which found   a long--bridge position as the  adsorption site.
However, we find that
the site of lowest total energy, i.e., the actual  adsorption site, is the  short--bridge
 position ${\bf A_{\rm 4}}$ (Fig.\ \ref{fig:pes}(b)).
If the adatom is adsorbed  at this position the surface dimer is broken and  the adatom 
forms  directional bonds with  the As atoms of the broken dimer.
Such  directional bonds are not   formed 
when the adatom is located at one of the  long--bridge sites ${\bf A_{\rm 1}}$ or  ${\bf A_{\rm 2}}$
where the adatom does  not  break any  surface bond. Therefore, the adatom can interact at an 
${\bf A_{\rm 1}}$ or  ${\bf A_{\rm 2}}$
site only with the completely
filled dangling bonds of the As--dimers. Accordingly the adsorption energies at the  long--bridge  sites
are by about 0.6\,eV smaller than at the short--bridge  sites (${\bf A_{\rm 3}}$, ${\bf A_{\rm 4}}$).  

In order to  obtain the   diffusivity 
the  jump rates on both sheets of the PES between the various sites of local total energy minima
were combined
by using a   random walk model  on a periodic lattice with inequivalent sites.\cite{ree:60}
The resulting mobilities along the [110] and [$\overline{\rm 1}$10] direction
are  $D_{\rm [\overline{\rm 1}10]}=$\,0.015\,cm$^2$/s\ exp($-$1.2\,eV/$k_{\rm B}T$ )
and  $D_{\rm [110]}=$\,0.03\,cm$^2$/s\ exp($-$1.5\,eV/$k_{\rm B}T$) for Ga adatoms and
\,0.03\,cm$^2$/s\ exp($-$1.4\,eV/$k_{\rm B}T$)
and 
0.05\,cm$^2$/s\ exp($-$1.6\,eV/$k_{\rm B}T$) for Al, respectively.
These diffusion constants are significantly smaller  than  previously published values
determined by {\it ab initio} total energy calculations.\cite{ohno:94} 
This difference is due to the fact that 
the activation energy of diffusion is mainly determined by the energy needed to jump
away from the adsorption site, which was not correctly identified  in former investigations.

The calculated diffusion constants exhibit a pronounced anisotropy.
As  it is  already obvious in Fig.\ \ref{fig:pes},
the  barriers for diffusion between two adjacent ${\bf A_{\rm 4}}$ sites
   are  smaller along the
[$\overline{\rm 1}$10] direction than along the  [110] direction.
Accordingly,  diffusion is fastest along the [$\overline{\rm 1}$10] direction.
Comparing the PES of Ga and Al adatoms we find no qualitative differences.
But the PES for an Al adatom 
is more strongly corrugated and therefore it  has higher diffusion barriers. This is a consequence
of  the larger  adatom--As bond strength for Al as reflected by the higher 
cohesive energy of AlAs compared to that of GaAs. Therefore, Al diffuses slower than
Ga in spite of the fact that  Al has a higher pre--exponential factor  (see Eq.\ ({\ref{eq:vorfaktor})), 
which is mainly caused by the lower mass of   Al as compared to  Ga.

\end{document}